\documentclass[twocolumn,showpacs,prl,floatfix]{revtex4}

\usepackage{graphicx}
\usepackage{bm}
\usepackage{psfrag}
\usepackage{amsmath}
\usepackage{amssymb}
\usepackage{latexsym}
\usepackage{exscale}

\newcommand{\vect}[1]{\bm{#1}}
\newcommand{\ten}[1]{\mbox{\textbf{{\textsf{#1}}}}}

\newcommand{\sprod}{\!\cdot\!}
\newcommand{\tprod}{}
\newcommand{\vprod}{\!\times\!}
\newcommand{\trace}{\operatorname{Tr}}

\newcommand{\dif}{\mathrm{d}}
\newcommand{\mi}{\mathrm{i}} 
\newcommand{\me}{\mathrm{e}}

\newcount\minute
\newcount\hour
\def\currenttime{%
    \minute\time
    \hour\minute
    \divide\hour60
    \the\hour:\multiply\hour60\advance\minute-\hour\the\minute}

\begin{document}

\title{Thermal Casimir vs Casimir--Polder forces: Equilibrium and
non-equilibrium forces}

\author{Stefan Yoshi Buhmann}
\author{Stefan Scheel}
\affiliation{Quantum Optics and Laser Science, Blackett Laboratory,
Imperial College London, Prince Consort Road,
London SW7 2AZ, United Kingdom}

\date{\today, \currenttime}

\begin{abstract}
We critically discuss whether and under what conditions Lifshitz
theory may be used to describe thermal Casimir--Polder forces on
atoms or molecules. An exact treatment of the atom--field coupling
reveals that for a ground-state atom (molecule), terms associated with
virtual-photon absorption lead to a deviation from the traditional
Lifshitz result; they are identified as a signature of non-equilibrium
dynamics. Even the equilibrium force on a thermalized atom (molecule)
may be overestimated when using the ground-state polarizability
instead of its thermal counterpart.
\end{abstract}

\maketitle


Dispersion forces such as Casimir and Casimir--Polder (CP) forces are
of increasing relevance in nanophysics \cite{0784}; recent successes
include the chemical identification of surface atoms via atomic force
microscopy \cite{0775} as well as the construction of novel biomimetic
dry adhesives \cite{0785}. They become important in efforts to
miniaturize atom chips \cite{0788} and play a key role in experiments
placing upper bounds on non-standard gravitational forces
\cite{0786}. 

In all of these areas, a thorough understanding of dispersion forces
under realistic conditions must account for their temperature
dependence induced by thermal photons \cite{0057}. A series of
high-precision experiments \cite{0575} has triggered a renewed
interest in this thermal Casimir force \cite{0621} by opening
the perspective of its experimental investigation \cite{0613}. It was
noticed that, depending on the model chosen to describe the metal
response, Lifshitz theory can yield different answers for the
temperature dependence of the Casimir energy between two metal plates
\cite{0610}. The resulting debate concerning the correct description
of the thermal Casimir force \cite{0681} will ultimately have to be
settled by experiments. Non-equilibrium situations of two plates of
different temperatures have recently been suggested as sensitive
probes to the quantum electrodynamics (QED) of the Casimir effect
\cite{0762}.

The CP force on single atoms can be measured indirectly via
spectroscopic means; clear evidence for thermal frequency shifts has
been found for atoms inside a cavity \cite{0145} and their signature
has been detected in the interaction of atoms with a sapphire surface
\cite{0769}. The novel non-equilibrium CP forces predicted for the
case of different surface and environment temperatures \cite{0394}
have recently been observed via their effect on the center-of-mass
oscillations of a trapped Bose--Einstein condensate \cite{0771}. While
some theoretical studies of the thermal CP force are based on a QED
treatment of the atom--field coupling \cite{0034,0768}, the vast
majority of investigations invokes a macroscopic calculation using
Lifshitz theory \cite{0057,0399,0182} or a linear-response description
of the atom \cite{0394,0037} (leading to equivalent results).

The macroscopic approach to the CP force is based on a very close
relation between Casimir and CP forces. It is the validity of this
one-to-one correspondence and its results for the CP force that we
intend to investigate in this Letter, so let us briefly recount the
argument in its traditional form: Generalizing the famous Lifshitz
result for two plates \cite{0057}, the thermal Casimir force on a
homogeneous body of arbitrary shape occupying a volume $V$ in free
space due to the presence of another body can be given as a Matsubara
sum \cite{0663,0696}
\begin{align}
\label{Eq0}
&\vect{F}=2k_\mathrm{B}T\int_{V}\dif^3r
 \sum_{N=0}^\infty\bigl(1-{\textstyle\frac{1}{2}}\delta_{N0}\bigr)
 \nonumber\\
&\times\Bigl[\trace\Bigl\{\ten{I}\vprod
 \bigl[\bm{\nabla}\vprod\bm{\nabla}\vprod\,
 +(\xi_N/c)^2\bigr]
 \ten{G}^{(1)}(\vect{r},\vect{r}',\mi\xi_N)\vprod
 \overleftarrow{\bm{\nabla}}'\Bigr\}_{\vect{r}=\vect{r}'}
 \nonumber\\
&\qquad-(\xi_N/c)^2\bm{\nabla}\sprod
 \ten{G}^{(1)}(\vect{r},\vect{r},\mi\xi_N)\Bigr]
\end{align}
with Matsubara frequencies $\xi_N=2\pi k_\mathrm{B}TN/\hbar$. Here,
$\ten{G}^{(1)}$ is the scattering part of the classical Green tensor
\begin{equation}
\label{Eq4}
\bigl[\bm{\nabla}\!\times\!\bm{\nabla}\!\times
 \,-\,(\omega/c)^2\,\varepsilon(\vect{r},\omega)\bigr]
 \ten{G}(\vect{r},\vect{r}',\omega)
 =\bm{\delta}(\vect{r}-\vect{r}')
\end{equation}
associated with the bodies that are characterized by their dielectric
permittivity $\varepsilon(\vect{r},\omega)$. The result~(\ref{Eq0})
can be obtained from the zero-temperature force by applying the simple
replacement rule
\begin{equation}
\label{Eq0b}
\frac{\hbar}{\pi}\int_{0}^{\infty}\dif\xi\,f(\mi\xi)\
 \mapsto\ 2k_\mathrm{B}T\sum_{N=0}^\infty
 \bigl(1-{\textstyle\frac{1}{2}}\delta_{N0}\bigr)f(\mi\xi_N)
\end{equation}
which is equivalent to replacing the zero-point energy
$\frac{1}{2}\hbar\omega$ by a thermal spectrum
$\bigl[n(\omega)\!+\!\frac{1}{2}\bigr]\hbar\omega$ with photon
number
$n(\omega)\!=\!1/\bigl[\me^{\hbar\omega/(k_\mathrm{B}T)}\!-\!1\bigr]$.
To derive the CP force, one assumes that one body consists of a dilute
gas (number density $\eta$) of atoms with polarizability
$\alpha(\omega)$, so that its permittivity may be approximated by 
$\varepsilon(\omega)=1+\eta\alpha(\omega)/\varepsilon_0$. After a
linear expansion in $\alpha$, it follows that the thermal CP
force is given by \cite{0663,0696}
\begin{multline}
\label{Eq0c}
\vect{F}(\vect{r}_{\!A})
 =-\mu_0k_\mathrm{B}T\sum_{N=0}^\infty
 \bigl(1-{\textstyle\frac{1}{2}}\delta_{N0}\bigr)\xi_N^2
 \alpha(\mi\xi_N)\\
\times\vect{\nabla}_{\!\!A}\trace 
 \ten{G}^{(1)}(\vect{r}_{\!A},\vect{r}_{\!A},\mi\xi_N).
\end{multline}
As an immediate consequence of its macroscopic derivation, this force
expression is very similar in structure to the respective Casimir
force. In particular, the replacement rule~(\ref{Eq0b}) is again valid
which often serves as a starting point for studies of the thermal CP
force \cite{0359}.

In view of the high precision of modern experiments, the current
widespread use of the macroscopic result~(\ref{Eq0c}) and the
replacement rule~(\ref{Eq0b}) for calculating thermal CP forces calls
for a critical discussion of their validity. In this Letter we clarify
the extent to which Eq.~(\ref{Eq0c}) can be used on the basis of a
direct calculation of the time-dependent thermal CP force from the
atom--field coupling Hamiltonian. We will demonstrate that force
components associated with thermal-photon absorption will generically
lead to deviations from Eq.~(\ref{Eq0c}). As we will show, the
macroscopic result provides a reasonable approximation to the force on
a fully thermalized atom, provided that the atomic polarizability is
correctly interpreted.

In electric dipole approximation, the force on an atom or a molecule 
(called atomic system in the following) prepared in an incoherent
superposition of internal eigenstates $|n\rangle$ is given by
\cite{0008,0696}
\begin{equation}
\label{Eq1}
\vect{F}(\vect{r}_{\!A},t)
=\bigl\langle\bigl[\vect{\nabla}\hat{\vect{d}}
 \sprod\hat{\vect{E}}(\vect{r})\bigr]_{\vect{r}=\vect{r}_{\!A}}
 \bigr\rangle,
\end{equation}
where it is assumed that the internal and translational motion of the
atomic system decouple in the spirit of Born--Oppenheimer
approximation. In order to evaluate this expression, one needs to
solve the coupled dynamics of the electromagnetic field and the atomic
system, which is governed by the Hamiltonian \cite{0003}
\begin{multline}
\label{Eq2}
\hat{H}=\int\dif^3r\int_0^\infty\dif\omega\,\hbar\omega\,
 \hat{\vect{f}}^\dagger(\vect{r},\omega)
 \sprod\hat{\vect{f}}(\vect{r},\omega)\\
+\sum_n E_n \hat{A}_{nn}
 -\sum_{m,n}\hat{A}_{mn}\vect{d}_{mn}\sprod
 \hat{\vect{E}}(\vect{r}_{\!A})
\end{multline}
($\vect{d}_{mn}=\langle m|\hat{\vect{d}}|n\rangle$,
$\hat{A}_{mn}=|m\rangle\langle n|$). The bosonic dynamical
variables $\hat{\vect{f}}^\dagger$ and $\hat{\vect{f}}$, which
describe the elementary excitations of the body-assisted
electromagnetic field, can be used to construct an expansion of the
electric-field operator $\hat{\vect{E}}(\vect{r})=%
\int_0^{\infty}\!\dif\omega\,%
\underline{\hat{\vect{E}}}(\vect{r},\omega)+\operatorname{H.c.}$
according to
\begin{multline}
\label{Eq3}
 \underline{\hat{\vect{E}}}(\vect{r},\omega)
=\mi\,\sqrt{\hbar/(\pi\varepsilon_0)}\,(\omega/c)^2
 \int\dif^3r'\,\sqrt{\operatorname{Im}\varepsilon(\vect{r}',\omega)}\\
 \times\ten{G}(\vect{r},\vect{r}',\omega)
 \!\cdot\!\hat{\vect{f}}(\vect{r}',\omega).
\end{multline}
The expansion coefficients are given in terms of the classical Green
tensor~(\ref{Eq4}) which obeys the integral relation
\begin{multline}
\label{Eq5}
\frac{\omega^2}{c^2}\int\!\dif^3s\,
 \operatorname{Im}\,\varepsilon(\vect{s},\omega)
 \,\ten{G}(\vect{r},\vect{s},\omega)
 \sprod\ten{G}^\ast(\vect{s},\vect{r}',\omega)\\
 =\mathrm{Im}\,\ten{G}(\vect{r},\vect{r}',\omega).
\end{multline}

The Hamiltonian~(\ref{Eq2}) implies that the atom--field dynamics is
governed by the system of equations
\begin{multline}
\label{Eq6}
\dot{\hat{A}}_{mn}=\mi\omega_{mn}\hat{A}_{mn}\\
 +\frac{\mi}{\hbar}\sum_k
 \bigl(\vect{d}_{nk}\hat{A}_{mk}-\vect{d}_{km}\hat{A}_{kn}\bigr)\sprod
 \hat{\vect{E}}(\vect{r}_{\!A}),
\end{multline}
\vspace*{-2ex}
\begin{multline}
\label{Eq7}
\dot{\hat{\!\vect{f}}}(\vect{r},\omega)
 =-\mi\omega\hat{\vect{f}}(\vect{r},\omega)
 +\sqrt{\operatorname{Im}\varepsilon(\vect{r},\omega)/
 (\hbar\pi\varepsilon_0)}\,(\omega/c)^2\\
\times\sum_{m,n}\hat{A}_{mn}\vect{d}_{mn}\sprod
 \ten{G}^\ast(\vect{r}_{\!A},\vect{r},\omega).
\end{multline}
We eliminate the field by formally integrating Eq.~(\ref{Eq7}) and
substituting the result back into Eq.~(\ref{Eq6}), which we arrange in
normal ordering. After using the integral relation~(\ref{Eq5}), we
obtain
\begin{align}
\label{Eq8}
&\dot{\!\hat{A}}_{mn}(t)
 =\mi\omega_{mn} \hat{A}_{mn}(t)
 +\hat{Z}_{mn}(t)\nonumber\\
&+\frac{\mi}{\hbar}\sum_k\int_0^\infty\!\!\!\dif\omega
 \bigl\{\me^{\mi\omega t}
 \underline{\hat{\vect{E}}}{}^\dagger(\vect{r}_{\!A},\omega)\sprod\!
 \bigl[\vect{d}_{nk}\hat{A}_{mk}(t)-\vect{d}_{km}\hat{A}_{kn}(t)\bigr]
 \nonumber\\
&+\me^{-\mi\omega t}\bigl[\hat{A}_{mk}(t)\vect{d}_{nk}
-\hat{A}_{kn}(t)\vect{d}_{km}\bigr]\sprod
 \underline{\hat{\vect{E}}}(\vect{r}_{\!A},\omega)\bigr\}
\end{align}
where
\begin{multline}
\label{Eq9}
\hat{Z}_{mn}(t)
=\frac{\mu_0}{\hbar\pi}
 \sum_{j,k,l}\int_0^\infty\dif\omega\,\omega^2
 \vect{d}_{kl}\sprod\operatorname{Im}
 \ten{G}(\vect{r}_{\!A},\vect{r}_{\!A},\omega)
 \int_0^t\dif\tau\\
\sprod\bigl\{\me^{-\mi\omega(t-\tau)}\bigl[
 \vect{d}_{jm}\hat{A}_{jn}(t)-\vect{d}_{nj}\hat{A}_{mj}(t)\bigr]
 \hat{A}_{kl}(\tau)\\
+\me^{\mi\omega(t-\tau)}\hat{A}_{kl}(\tau)\bigl[
 \vect{d}_{nj}\hat{A}_{mj}(t)-\vect{d}_{jm}\hat{A}_{jn}(t)\bigr]
 \bigr\}
\end{multline}
denotes the zero-point contribution to the internal atomic dynamics.
The thermal contribution can be determined iteratively by substituting
the self-consistent solution
\begin{align}
\label{Eq10}
&\hat{A}_{mn}(t)=\me^{\mi\tilde{\omega}_{mn}t}\hat{A}_{mn}
 +\frac{\mi}{\hbar}\sum_k\int_0^\infty\dif\omega
 \int_0^t\dif\tau\,\me^{\mi\tilde{\omega}_{mn}(t-\tau)}
 \nonumber\\
&\times\bigl[\hat{A}_{mk}(\tau)\vect{d}_{nk}
 -\hat{A}_{kn}(\tau)\vect{d}_{km}\bigr]\sprod
\bigl[\underline{\hat{\vect{E}}}(\vect{r}_{\!A},\omega)
 \me^{-\mi\omega\tau}+\operatorname{H.c.}\bigr]
\end{align}
into the truncated Eq.~(\ref{Eq8}) [without the zero-point
contribution $\hat{Z}_{mn}(t)$] back into itself and taking
expectation values. With the field initially being prepared
in a thermal state $\hat{\rho}_T=%
\exp[-\hat{H}_\mathrm{F}/(k_\mathrm{B}T)]/%
\trace\{\exp[-\hat{H}_\mathrm{F}/(k_\mathrm{B}T)]\}$,
$\hat{H}_\mathrm{F}=\int\dif^3r\int_0^\infty\dif\omega\,\hbar\omega%
\hat{\vect{f}}^\dagger(\vect{r},\omega)\sprod%
\hat{\vect{f}}(\vect{r},\omega)$,
the non-vanishing averages of the field operators are
\begin{multline}
\label{Eq11}
\bigl\langle\underline{\hat{\vect{E}}}{}^\dagger(\vect{r},\omega)
\underline{\hat{\vect{E}}}(\vect{r}',\omega')\bigr\rangle\\
=(\hbar\mu_0/\pi)n(\omega)\omega^2
 \operatorname{Im}\ten{G}(\vect{r},\vect{r}',\omega)
 \delta(\omega-\omega')
\end{multline}
[recall Eqs.~(\ref{Eq3}) and (\ref{Eq5})]. One thus obtains a closed
system of equations for the internal atomic dynamics
\begin{equation}
\label{Eq12}
\bigl\langle\dot{\hat{A}}_{mn}\bigl\rangle
 =\mi\omega_{mn}\bigl\langle\hat{A}_{mn}\bigl\rangle
 +\bigl\langle\hat{Z}_{mn}\bigl\rangle
 +\bigl\langle\hat{T}_{mn}\bigl\rangle,
\end{equation}
with the thermal contribution being given by
\begin{multline}
\label{Eq13}
\hspace{-2ex}\bigl\langle\hat{T}_{mn}(t)\bigl\rangle\\
=\frac{\mu_0}{\hbar\pi}
 \sum_{k,l}\int_0^\infty\dif\omega\,\omega^2n(\omega)\int_0^t\dif\tau
 \bigl[\me^{-\mi\omega(t-\tau)}+\me^{\mi\omega(t-\tau)}\bigr] \\ \times
\bigl\{ \me^{\mi\tilde{\omega}_{mk}(t-\tau)}
\vect{d}_{nk}\sprod\operatorname{Im}
 \ten{G}(\vect{r}_{\!A},\vect{r}_{\!A},\omega) \\
 \sprod\bigl[\vect{d}_{lm}\bigl\langle\hat{A}_{lk}(\tau)\bigl\rangle
 -\vect{d}_{kl}\bigl\langle\hat{A}_{ml}(\tau)\bigl\rangle\bigr]\\
+\me^{\mi\tilde{\omega}_{kn}(t-\tau)}
 \vect{d}_{km}\sprod\operatorname{Im}
 \ten{G}(\vect{r}_{\!A},\vect{r}_{\!A},\omega) \\
 \sprod\bigl[\vect{d}_{nl}\bigl\langle\hat{A}_{kl}(\tau)\bigl\rangle
 -\vect{d}_{lk}\bigl\langle\hat{A}_{ln}(\tau)\bigl\rangle
 \bigr]\bigr\}.
\end{multline}
For weak atom--field coupling, this system can be solved in Markov
approximation using the relations
$\bigl\langle\hat{A}_{mn}(\tau)\bigr\rangle\simeq%
\me^{-\mi\tilde{\omega}_{mn}(t-\tau)}%
\bigl\langle\hat{A}_{mn}(t)\bigr\rangle$ and
$\int_0^t\dif\tau\,\me^{\mi x(t-\tau)}\simeq\pi\delta(x)%
+\mi\mathcal{P}/x$.

For a non-degenerate system, the off-diagonal elements of the atomic
density matrix $\hat{\sigma}$ decouple from one another as well as
from the diagonal elements, and one finds that the internal atomic
dynamics follows the rate equations
\begin{eqnarray}
\label{Eq14}
\dot{\sigma}_{nn}(t)
&\!=&\!-\Gamma_n\sigma_{nn}(t)
 +\sum_k\Gamma_{kn}\sigma_{kk}(t),\\
\label{Eq15}
\dot{\sigma}_{mn}(t)
&\!=&\!\bigl[-\mi\tilde{\omega}_{mn}
 -{\textstyle\frac{1}{2}}(\Gamma_m+\Gamma_n)\bigr]
 \sigma_{mn}(t),\,m\neq n\quad
\end{eqnarray}
($\sigma_{mn}\!=\!\langle m|\hat{\sigma}|n\rangle\!=%
\!\bigl\langle\hat{A}_{nm}\bigr\rangle$). Here, the total loss rates
read 
\begin{multline}
\label{Eq16}
\hspace{-2ex}\Gamma_n=\sum_k\Gamma_{nk}
=\frac{2\mu_0}{\hbar}\!\sum_k
 \tilde{\omega}_{nk}^2
 \{\Theta(\tilde{\omega}_{nk})[n(\tilde{\omega}_{nk})+1]\\
+\Theta(\tilde{\omega}_{kn})n(\tilde{\omega}_{kn})\}
 \vect{d}_{nk}\sprod\operatorname{Im}
 \ten{G}(\vect{r}_{\!A},\vect{r}_{\!A},|\tilde{\omega}_{nk}|)
 \sprod\vect{d}_{kn}
\end{multline}
and the shifts of the atomic transition frequencies
$\tilde{\omega}_{mn}=\omega_{mn}+\delta\omega_m-\delta\omega_n$ 
are given by
\begin{multline} 
\label{Eq17}
\delta\omega_n=\sum_k \delta\omega_{nk}
=\frac{\mu_0}{\pi\hbar}\sum_k
 \mathcal{P}\int_0^\infty\dif\omega\,\omega^2
 \biggl[\frac{n(\omega)+1}{\tilde{\omega}_{nk}-\omega}\\
+\frac{n(\omega)}{\tilde{\omega}_{nk}+\omega}\biggr]
 \vect{d}_{nk}\sprod\operatorname{Im}
 \ten{G}(\vect{r}_{\!A},\vect{r}_{\!A},\omega)\sprod
 \vect{d}_{kn};
\end{multline}
they reduce to the perturbative result derived in Ref.~\cite{0768}
upon setting $\tilde{\omega}_{nk}\simeq\omega_{nk}$ in the
denominators.

With these preparations, we evaluate the CP force~(\ref{Eq1})
with the help of Eqs.~(\ref{Eq3}), (\ref{Eq10}) and (\ref{Eq11}) and
the solution to Eq.~(\ref{Eq7}) as
\begin{align} 
\label{Eq18}
&\vect{F}(\vect{r}_{\!A},t)\nonumber\\
&=\frac{\mi\mu_0}{\pi}\sum_{n,k}\int_0^\infty\dif\omega\,\omega^2
 \vect{\nabla}\tprod\vect{d}_{nk}\sprod
 \mathrm{Im}\,\ten{G}^{(1)}(\vect{r},\vect{r}_{\!A},\omega)
 \sprod\vect{d}_{kn}\bigr|_{\vect{r}=\vect{r}_{\!A}}\nonumber\\
&\times\int_0^t\dif\tau\,\bigl\langle\hat{A}_{nn}(\tau)\bigl\rangle
 \bigl\{n(\omega)
 \me^{[\mi(\omega+\omega_{nk})-(\Gamma_n+\Gamma_k)/2](t-\tau)}
 \nonumber\\
&+[n(\omega)+1]
 \me^{[-\mi(\omega-\omega_{nk})-(\Gamma_n+\Gamma_k)/2](t-\tau)}\bigr\}
 +\operatorname{c.c.}
\end{align}
Here we have used the correlation function
\begin{equation}
\label{Eq19}
\bigl\langle\hat{A}_{mn}(t)\hat{A}_{kl}(\tau)\bigr\rangle
 =\delta_{nk}\bigl\langle\hat{A}_{ml}(\tau)\bigr\rangle
 \me^{[\mi\tilde{\omega}_{mn}-(\Gamma_m+\Gamma_n)/2](t-\tau)}
\end{equation}
which follows from Eq.~(\ref{Eq15}) by means of the quantum regression
theorem \cite{0605}. Evaluating the $\tau$ integral in Markov
approximation and the $\omega$ integral by means of contour-integral
techniques (cf. Ref.~\cite{0008}), one finds that the thermal CP force
on an atomic system prepared in an incoherent superposition of energy
eigenstates is given by
$\vect{F}(\vect{r}_{\!A},t)=\sum_{n}\sigma_{nn}(t)\vect{F}_n(\vect{r}_{\!A})$
with force components
\begin{widetext}
\begin{multline}
\label{Eq20}
\vect{F}_n(\vect{r}_{\!A})
 =-\mu_0k_\mathrm{B}T\sum_{N=0}^\infty
 \bigl(1-{\textstyle\frac{1}{2}}\delta_{N0}\bigr)\xi_N^2
 \vect{\nabla}\trace\bigl\{
 [\bm{\alpha}_n(\mi\xi_N)
 +\bm{\alpha}_n(-\mi\xi_N)]
 \sprod\ten{G}^{(1)}(\vect{r}_{\!A},\vect{r},\mi\xi_N)
 \bigr\}\bigr|_{\vect{r}=\vect{r}_{\!A}}
+\mu_0\sum_k
\bigl\{\Theta(\tilde{\omega}_{nk})
 \Omega^2_{nk} \\ 
\times[n(\Omega_{nk})+1]
 \vect{\nabla}\vect{d}_{nk}\sprod
 \ten{G}^{(1)}(\vect{r},\vect{r}_{\!A},\Omega_{nk})
 \sprod\vect{d}_{kn}\bigr|_{\vect{r}=\vect{r}_{\!A}}
 -\Theta(\tilde{\omega}_{kn})\Omega^{\ast 2}_{kn}n(\Omega_{kn}^\ast)
 \vect{\nabla}\vect{d}_{nk}\sprod
 \ten{G}^{(1)}(\vect{r},\vect{r}_{\!A},\Omega_{kn}^\ast)
 \sprod\vect{d}_{kn}\bigr|_{\vect{r}=\vect{r}_{\!A}}
 +\mathrm{c.c.}\bigr\}
\end{multline}
\end{widetext}
[$\Omega_{nk}=\tilde{\omega}_{nk}+\mi(\Gamma_n+\Gamma_k)/2$] and
atomic/molecular polarizability
\begin{equation}
\label{Eq21}
\bm{\alpha}_n(\omega)
=\frac{1}{\hbar}\sum_k\biggl[
 \frac{\vect{d}_{nk}\tprod\vect{d}_{kn}}
 {-\Omega_{nk}-\omega}
 +\frac{\vect{d}_{kn}\tprod\vect{d}_{nk}}
 {-\Omega_{nk}^\ast+\omega}
 \biggr].
\end{equation}
This result generalizes the zero-temperature force calculated in
Ref.~\cite{0008}.

In order to compare this force with the macroscopic
result~(\ref{Eq0c}), we consider an isotropic atomic system in the
perturbative limit $\Omega_{nk}\simeq\omega_{nk}$, whereby
Eq.~(\ref{Eq20}) simplifies to
\begin{multline}
\label{Eq22}
\hspace{-2ex}\vect{F}_n(\vect{r}_{\!A})
=-\mu_0k_\mathrm{B}T\sum_{N=0}^\infty
 \bigl(1-{\textstyle\frac{1}{2}}\delta_{N0}\bigr)
\\ \times
\xi_N^2
 \alpha_n(\mi\xi_N)\vect{\nabla}_{\!\!A}\trace 
 \ten{G}^{(1)}(\vect{r}_{\!A},\vect{r}_{\!A},\mi\xi_N)\\
+\frac{\mu_0}{3}\sum_k\omega_{nk}^2
 \{\Theta(\omega_{nk})[n(\omega_{nk})+1]
 -\Theta(\omega_{kn})n(\omega_{kn})\}\\
\times|\vect{d}_{nk}|^2\vect{\nabla}_{\!\!A}\trace\operatorname{Re} 
 \ten{G}^{(1)}(\vect{r}_{\!A},\vect{r}_{\!A},\omega_{nk})
\end{multline}
and coincides with the negative gradient of the frequency
shift~(\ref{Eq17}) given in Ref.~\cite{0768}. In addition to the
Matsubara sum, the thermal CP force has resonant contributions
proportional to $n(\omega_{nk})$ and $n(\omega_{nk})\!+\!1$,
respectively, which are due to the absorption and emission of
photons by the atomic system. Even the force
$\vect{F}_0(\vect{r}_{\!A})$ on a
ground-state atom or molecule exhibits resonant force components
$-\frac{1}{3}\mu_0\omega_{k0}^2n(\omega_{k0})|\vect{d}_{0k}|^2%
\vect{\nabla}_{\!\!A}\trace\operatorname{Re}%
\ten{G}^{(1)}(\vect{r}_{\!A},\vect{r}_{\!A},\omega_{k0})$
associated with thermal-photon absorption which lead to a discrepancy
with the macroscopic result~(\ref{Eq0c}). These resonant forces can be
observed on time scales which are short with respect to the inverse
ground-state heating rates $\Gamma_{0k}^{-1}$, and their magnitude
scales with the number of thermal photons at the respective transition
frequency $n(\omega_{k0})$. Polar molecules thus present an ideal
candidate for studying them, since their heating time can be of the
order of several seconds, and the thermal photon number at the
relevant vibrational and rotational transition frequencies in the
microwave regime can reach values of up to a few hundred at room
temperature \cite{0772}. As discussed in Ref.~\cite{0769}, the
enhanced frequency shifts observed very recently via selective
reflection spectroscopy of cesium in the far infrared might be due to
a resonant thermal effect of this kind.

Our approach allows us to discuss the full dynamics of the CP force.
It is thus able to reveal that resonant force components associated
with absorption and stimulated / spontaneous emission of photons are a
genuine non-equilibrium effect: According to Eq.~(\ref{Eq14}), the
atomic system reaches thermal equilibrium with its environment in the
long-time limit and is described by the thermal state $\hat{\sigma}_T%
=\me^{-\sum_n\tilde{E}_n\hat{A}_{nn}/(k_\mathrm{B}T)}/\trace\bigl[%
\me^{-\sum_n\tilde{E}_n\hat{A}_{nn} /(k_\mathrm{B}T)}\bigr]$
($\tilde{E}_n=E_n+\hbar\delta\omega_n$). In thermal equilibrium, all
resonant force components cancel and the force can be written in the
form of Eq.~(\ref{Eq0c}) only if the atomic polarizability has been
identified as the thermal polarizability, 
$\alpha_T(\omega)=\sum_n\sigma_{T,nn}\alpha_n(\omega)$. In the limit
of a single dominant transition, this equilibrium force is smaller
than that resulting from Eq.~(\ref{Eq0c}) with $\alpha_0(\omega)$ by a
factor $r_T$ $\!=$
$\!\tanh[\hbar\tilde{\omega}_{10}/(2k_\mathrm{B}T)]%
=[2n(\tilde{\omega}_{10})+1]^{-1}$. The force on ground-state atoms,
which is dominated by electronic transitions, is insensitive to this
effect at room temperature due to the very small thermal photon
number: For Rb (\mbox{$\omega_{10}$ $\!=$
$\!2.37\!\times\!10^{15}\mbox{Hz rad}$} \cite{0822}), one obtains
\mbox{$r_T$ $\!=$ $\!1-1.3\!\times\!10^{-26}$} at \mbox{$T$ $\!=$
$\!300\mathrm{K}$}. As in the case of the non-equilibrium force
discussed above, a noticeable deviation from the Lifshitz result is to
be expected for polar molecules: For CaF, force components associated
with vibrational transitions (\mbox{$\omega_{10}$ $\!=$
$\!1.15\!\times\!10^{14}\mbox{Hz rad}$} \cite{0789}) are reduced by a
factor \mbox{$r_T$ $\!=$ $\!0.90$}, those associated with rotational
transitions (\mbox{$\omega_{10}$ $\!=$
$\!1.32\!\times\!10^{11}\mbox{Hz rad}$} \cite{0819}) even by a factor 
\mbox{$r_T$ $\!=$ $\!0.0017$}---they are thus strongly reduced with
respect to the prediction from Lifshitz theory.

In conclusion, a full quantum-mechanical treatment of the atom--field
interaction has revealed that the thermal CP force on a ground-state
atom or molecule cannot be obtained from a macroscopic calculation
while the force on a fully thermalized atom can---provided that one
uses its correct finite-temperature polarizability. As shown, the
force on thermalized molecules is considerably smaller than suggested
by Lifshitz theory. The resonant contributions to the force that occur
for a ground-state atom are due to the absorption of thermal photons
and manifest a non-equilibrium effect. In particular, they may
dominate the thermal CP force on polar molecules, where they present
both a considerable limiting factor for the miniaturisation of
molecular surface traps and a novel probe to the surface-assisted
thermal dynamics of these molecules. The discussion presented here
immediately generalizes to scenarios such as those suggested in
Ref.~\cite{0394} where different parts of the environment are held at
different temperatures.


\acknowledgments
This work was supported by the Alexander von Humboldt Foundation and
the UK Engineering and Physical Sciences Research Council.


\end{document}